\documentclass{article}
\usepackage{spconf}
\title{WAVENET BASED LOW RATE SPEECH CODING}

\name{\begin{tabular}{c}
W. Bastiaan Kleijn,$^{1,3}$
Felicia S. C. Lim,$^{1}$
Alejandro Luebs,$^{1}$
 Jan Skoglund,$^{1}$\\
Florian Stimberg,$^{2}$
Quan Wang,$^{1}$
Thomas C. Walters$^{2}$
\end{tabular}
}

\address{$^{1}$Google Inc., San Francisco, CA; \hspace{0.5em}
$^{2}$ DeepMind, London, UK; \hspace{0.5em}
$^{3}$Victoria University of  Wellington, NZ\\
}

\usepackage{,amsmath,graphicx,url,times}
\usepackage{amsfonts}
\usepackage{xcolor}
\usepackage{algorithm,algpseudocode}
\usepackage{subcaption}
\graphicspath{{figures/}}

\def\A{\mathcal{A}}

\begin{document}
\ninept
\maketitle

\begin{abstract}
 Traditional parametric coding of speech facilitates low rate but provides poor
 reconstruction quality because of the inadequacy of the model used. We describe how a
 WaveNet generative speech
 model can be used to generate high quality speech from the bit stream of a standard
 parametric coder operating at 2.4 kb/s. We compare this parametric
 coder with a waveform coder based on the same generative model and show that
 approximating the signal waveform incurs a large rate penalty. Our experiments confirm
 the high performance of the WaveNet based coder and show that the speech produced by the
 system is able to additionally perform implicit bandwidth extension and does not significantly
 impair recognition of the original speaker for the human listener, even when that speaker has
 not been used during the training of the generative model.
\end{abstract}

\begin{keywords}
Speech coding, parametric coding, WaveNet, generative model
\end{keywords}

\section {Introduction}
\label{s:introduction}

Speech coding found its first major application in secure
communications, e.g., \cite{tremain:76}, and later enabled low-cost mobile and
internet communications, e.g., \cite{Kroon:86,Salami94toll,Kleijn94Interpolation}. With
the continuously decreasing cost of bandwidth in most applications,
the trade-off between rate and quality has gravitated to higher rates,
typically over 16 kb/s, to ensure good quality. This state-of-the-art
may change with the advent of a new generation of coders that
provides a significant leap in performance. In this paper we discuss an approach
that can provide good quality at rates around 2-3 kb/s with significant
potential for further improvement in the rate-quality trade-off.

The redundancy in rate of existing speech coders can
be determined from estimates of the information rate in speech. A recent
rate estimate \cite{vanKuyk2016OnThe} based on comparing signals with the
same message is consistent with lexical information rates computed from phoneme statistics
\cite{denes1963staspoEng}.
They suggest that the true information rate is less than 100 b/s.
Attributes of speech that identify the speaker and speaking
style do not vary rapidly over time and hence do not change this
rough estimate significantly. The common coding algorithms used in current
communication systems require a rate that is roughly two orders of magnitude
higher than the rate of the information conveyed.

Essentially, all speech coding methods are based on an explicit model
of the signal, which usually is time varying. In \textit{parametric
 coding} the signal generation is generated at the decoder based on the %time-varying
model parameters only. The quality of the signal reproduced by parametric
coders is limited by the efficacy of the
model. However, even poor signal models can be usefully exploited for
high-quality coding. \textit{Waveform coding} exploits that conditioning
information (the model and its parameters) reduces the minimum rate
required to achieve a particular mean error for the signal waveform. The
penalty paid is that the reproduced signal is an
approximation of the original waveform. This requires the transmission
of information that, at least in principle, is not needed for high
fidelity reconstruction, explaining in part the high rate of current
speech coding schemes.

Most models used in speech coding have a statistical basis. A speech
signal can be described as a discrete-time stochastic process
$\{s_i\}$ with a non-zero (differential) entropy rate. A discrete-time
stochastic process can be characterized by a sequence of conditional
probability density functions (PDFs) $f(s_i |s_{i-1}, s_{i-2}, \cdots)$. If
the memory is $p$ samples, then the process is an order-$p$ Markov
processs. Application of the chain rule relates the PDF of a sample
sequence to that of the conditional PDFs.

Various generative models have been used to describe the speech
process. The models generally assume speech to be a Markov process and
provide a conditional distribution for the next signal sample given a
set of past samples. Ubiquitous are linear autoregressive (AR) models
\cite{Atal70Adaptive},
%itakura1970statistical}
and hidden Markov models (HMM) \cite{Tokuda:00}. Refined generative
models such as ARMA, e.g., \cite{Grenier83}, and kernel
density estimation (KDE)  HMM, e.g., \cite{piccardi2007hidden}, can also be
used. However, linear
AR modeling remains the most commonly used generative model in speech
coding. This may change with the recent introduction of the deep
neural network (DNN) based WaveNet \cite{WaveNet:2016a} generative
models.

Recursive sampling of the conditional
PDF of a speech model can be used to produce a speech
signal. The sampling of the PDF corresponds to the generation of new
information. To retain the perceptible attributes of an original speech signal in the coding
application, the PDF must include conditioning variables
(side information). These typically specify the
short-term power spectral density, pitch and periodicity
level. Good signal quality can be guaranteed even for imperfect
generative models by approximating the original signal
waveform. However, no new information is then generated during
reconstruction and this information must be transmitted instead. Efficiency can be
increased by encoding of blocks of samples simultaneously,
%\cite{gray2012source},
for example using analysis-by-synthesis \cite{bell1961reductio}
in the context of generative models. The analysis-by-synthesis paradigm applied to
AR models, introduced in \cite{Singhal1984}, is universally
%schroeder1985code},
used in mobile phone standards.

The contribution of this paper is the usage of WaveNet as a generic generative
model for speech coding and an analysis thereof. We describe two WaveNet based coding
architectures: \textit{i}) a parametric coder that encodes only the conditioning
variables, and \textit{ii}) a waveform coder that encodes the conditioning variables and
also the observed waveform exploiting the conditional distribution. The
parametric coder model differs from
\cite{Tamamori2017SpeakerdependentWV} in that it is not speaker dependent, and
that it can also be decoded with a conventional low complexity decoder. Our
coding architectures replace the text sequence used as conditioning information in the
original text-to-speech (TTS) application of WaveNet by a sequence of
quantized parameters of a parametric speech coder. This change is nontrivial
as in a coding scenario the generative model cannot be trained on the speakers
that the coder encounters during operation. It will be shown in section
\ref{s:experiments} that WaveNet can be generalized to the multi-speaker
case.

% Our approach to WaveNet based parametric coding has
%two advantages over approaches that use a DNN based analysis and
%quantization structure. First, as the the encoder is a conventional
%parametric speech coder it has low computational complexity. Second,
%the quality of the decoded signal depends only on the computational capacity
%of the decoder: it is possible to decode the signal with a low cost
%conventional decoder.

%We now briefly introduce approach \textit{ii}, which produces an
%approximated waveform as reconstruction. This method has the advantage that
%the  generative model does not have to be a good match for the signal
%to get good quality. Hence, if the training data for the DNNs
%model are not representative for the data that have to be encoded, the method
%can still produce high quality output. Approach \textit{ii} also can handle
%scenarios where the conditioning parameters are not correct. This
%makes this approach particularly relevant for the high-quality coding
%of music. Its drawbacks are that the encoder is computationally
%complex and that the encoder and decoder must run synchronously.

In the remainder of this paper, we first describe our approach in
more detail in section \ref{s:algorithm}, then discuss our experimental results in
section \ref{s:experiments} and finally provide conclusions in section
\ref{s:conclusions}.

%\tcb{ends top of page 2.}

\section {Algorithm}
\label{s:algorithm}
%\tcb{From top page 2 to middle page 3}

In this section we discuss the parametric coding architecture, analyze the
rate and describe the waveform coding architecture.

\subsection{Parametric WaveNet Coder}

A parametric coder transmits only the conditioning variables of the generative model that
generates the signal at the decoder. It is possible to train a neural
architecture at the encoder to optimize the conditioning
variables. In this paper, we opted for a conventional parametric encoder instead. The latter
approach has a significantly lower complexity at the encoder and
facilitates the use of a low complexity decoder if computational resources fall
short. We first motivate our conventional encoder choice and then describe the WaveNet decoder.

Traditional parametric coders almost always encode a similar set of parameters:
spectral envelope, pitch, and voicing level. The parameter
set differs little for approaches based on a temporal perspective with glottal pulse trains
\cite{atal1971speech,markel1976linear},  and those based on a frequency-domain perspective
with sinusoids \cite{Hedelin:81,McAulay:86}.  Any of these parameter sets can be used as
a set of conditioning variables for WaveNet.

To illustrate that our architecture does not require special features
for the encoder, we selected Codec
2 \cite{RoweWinNT}, an open source speech coder. It belongs to the sinusoidal
coder family and can run at various update rates. For example, at 2.4 kb/s, each 20 ms block encodes
the short-time spectral envelope using line spectral frequencies \cite{Itakura:75} with 36
bits, the pitch with 7 bits, the signal power with 5 bits, and the voicing level
with 2 bits. The voicing level is determined as in the
multiband excitation vocoder \cite{Griffin:88}.

% \tcb{In
% section \ref{s:experiments}  we discuss how the block update rate of these
% conditioning variables affects the performance of the WaveNet model.}

We note that parametric coders (including Codec 2) almost
universally operate on narrow-band speech with a sampling rate of 8~kHz
but for high quality output speech, a wide-band signal ($\ge$~16~kHz) is preferred.
Historically, wide-band extension is optionally applied after the decoder \cite{Iser2008}. In our approach,
we train our decoder with 8 kHz conditioning variables and 16 kHz
speech signals such that it implicitly performs bandwidth extension.

For the parametric decoder, we use the WaveNet generative
model \cite{WaveNet:2016a}. Given the past output signal and the conditioning variables,
WaveNet provides a discrete probability distribution of the next signal sample
using the 8-bit ITU-T G.711 $\mu$-law format. It then samples this
distribution to select the output sample value.

The WaveNet architecture is a multi-layer structure using dilated convolution with gated
cells. The conditional variables are supplied to all layers of the network. For the coder,
we retained the standard WaveNet configuration of \cite{WaveNet:2016a} but replaced the
conditioning variables with the decoded Codec 2 bit stream. The
Codec 2 decoder provides its parameters to its sinusoidal renderer at 100~Hz, which we
used unchanged as the conditioning variables for the WaveNet decoder. As WaveNet requires
conditioning for each output sample, we hold the conditional variables constant
for 10 ms intervals.

During training, WaveNet learns the parameters of a softmax function that represents the
conditional discrete probability distribution. The training is subject to the same
conditioning variables that are used  during run-time. In contrast to WaveNet training for
TTS applications, we used a training database containing a large
number of different talkers providing a wide variety of voice characteristics, all without
conditioning on a label that identifies the talker.

\subsection{Rate Analysis}
\label{s:rateanalysis}
The rate benefit of \textit{generating} the waveform over approximating the
original waveform can be estimated from the information rate generated at the decoder.
Basic information theory allows us to estimate the relevant rates. Let us
consider a generated signal sequence $\{S_i\}_{i\in\A}$, where $\A$  is an index sequence, and
a conditioning sequence $\{\Theta_i\}_{i\in\A}$, with the two sequences being presented at the
same sampling rate. Let both sequences be discrete valued and of finite
length $|\A|$. Their entropy rates then satisfy
\begin{align}
\frac{1}{|\A|} H(\{S_i\}, \{\Theta_i\}) =
 \frac{1}{|\A|}   H(\{S_i\}|\{\Theta_i\}) + \frac{1}{|\A|}  H(\{\Theta_i\}),
\end{align}
where we omitted sequence subscripts to simplify notation. Hence, the overall information
rate contained in the decoded signal is the sum of the information rate associated with the
generative process and the rate of the encoded conditioning parameters.
The rate required for the conditioning variables is upper bounded by
the rate of the parametric coder.

The information rate $ \frac{1}{|\A|}   H(\{S_i\}|\{\Theta_i\})$
associated with the generative process is simple to evaluate for
WaveNet. We consider $|\A| \rightarrow \infty$ and make the additional assumption that
speech is a short-time stationary and ergodic process and that, therefore, $\{S_i\}$ and $\{\Theta_i\}$ are
stationary.

\begin{sloppypar}
At its output, the WaveNet decoder produces a probability
distribution with a set $\mathcal{N}$ of 256 discrete values for a $\mu$-law encoding
of the next sample. We denote this distribution as $q^{(i)}_n$, $i \in
\mathbb{Z}$ , $n\in \mathcal{N}$. The distribution $q^{(i)}$ is
sampled to produce the scalar output signal value. The mean information in
bits generated by this sampling operation is the conditional entropy
of the distribution:   \begin{align}
H(S_i|s_{i-1},s_{i-2}, \cdots; \theta_i ) = - \sum_{n \in \mathcal{N}} q^{(i)}_n  \log_2 q^{(i)}_n ,
\end{align}
where $\theta_i$ is the current vector of conditioning variables, and where
capital letters indicate random variables and lower-case letters realizations. Note that
$H(S_i|s_{i-1},s_{i-2}, \cdots; \theta_i )$ is simple to
compute and that the evaluation is exact for the reconstruction process.
\end{sloppypar}

We can now find the generated signal rate using an approximation to
the chain rule:
\begin{align}
\lim_{|\A| \rightarrow \infty} \frac{1}{|\A|}  &  H(\{S_i\}|\{\Theta_i\})
=H(S_i|S_{i-1},S_{i-2}, \cdots; \Theta_i ) \\
 &\approx \frac{1}{|\A_0|} \sum_{i\in\A_0}H(S_i|s_{i-1},s_{i-2},
   \cdots; \theta_i ) ,
\label{q:genrate_est}
\end{align}
where $\A_0$ is an observed finite sequence. We will evaluate the
result in section \ref{s:experiments}.

\subsection{WaveNet Waveform Coder}
\label{s:waveformcoder}

As will be seen in section \ref{s:experiments}, relation
\eqref{q:genrate_est} implies that a coder that reproduces the signal
waveform is inefficient compared to a coder relying on a
generative model. However, a generative coder cannot guarantee its output quality;
situations may occur where the modelled generative
distribution is not a good description of the signal.

Importantly, scenarios where the parametric WaveNet coder does not perform well can be
detected within the WaveNet framework. Running a waveform WaveNet at the encoder allows
the evaluation of the log likelihood of the input signal based on the WaveNet generative
model and compare that to the expectation. Thus, we can select between parametric and waveform
coding (low and high rate). When generative performance is good, the decoder receives no
waveform information, and it reverts to a generative mode. Resynchronization can be achieved
with conventional techniques \cite{verhelst1993overlap}. We will report on this mode selection
paradigm elsewhere and only discuss the basic WaveNet waveform coder here.

%More-over, in these situations
%generally no accurate distortion measure is available  either. If such scenarios occur, it is
%advantageous to use waveform approximation and exploit an imperfect
%model to reduce the rate required to encode the signal. In a waveform coder,
%the waveform information is transmitted to the decoder instead of being generated
%at the decoder. In the present context,  this implies that we run
%WaveNet generators in synchrony at the encoder and the
%decoder.

Waveform coding is commonly based on prediction, particularly at rates below 30 kb/s.
While fixed-rate predictive coding is most common, variable-rate versions
also exist. In predictive coders, the conditional probability distribution is generally
considered fixed relative to its mean predicted value.
%Quantization and, if present, entropy coding is optimized based on this fixed conditional
%distribution.
The same approach can be used for WaveNet coding, but it is not natural. In WaveNet
waveform coding, the prediction step can be omitted as a conditional discrete distribution
of the next sample is available without actually computing a predicted sample value.

We now describe our variable-rate WaveNet waveform coding structure.
We first discuss the quantization step and the corresponding decoding step. Let $Q:
\mathbb{R} \rightarrow \mathcal{N}$ be the mapping from the signal
to its quantization index and $Z: \mathcal{N} \rightarrow
\mathbb{R}$ be the corresponding decoding operation. Then, a signal $x_i$ is encoded
at the encoder as $n_i = Q(x_i)$ and the quantized signal is obtained at the decoder as $\hat{x}_i = Z(n_i)$.
The resolution of $Q$ determines both the rate and the quality for the waveform coder. In
conventional WaveNet, $Q$ is a $\mu$-law quantizer. Our proposed approach
described below will therefore result in the exact same waveform as basic $\mu$-law coding.

In the WaveNet waveform coder, the sequence of quantization indices $\{n_i\}_{i\in\mathbb{Z}}$
is subject to entropy coding and subsequently transmitted over the channel along
with the conditioning variables.
Identically trained WaveNet models are required at both the encoder and decoder to provide
the conditional probability distribution $q^{(i)}$ to the entropy encoder and decoder.
Importantly, it is $\hat{x}_i$ that is used at the decoder as input for generating subsequent
samples in the WaveNet model, thus creating a \textit{closed loop} coder. (In contrast, the parametric
WaveNet coder takes the previous generated sample as input.)
Although $q^{(i)}$ is an approximate  distribution for the original signal
$\{x_i\}_{i\in\A}$, it can be used to reduce the rate significantly with entropy
coding. Using techniques such as arithmetic coding \cite{pasco:76, rissanen:79}
can provide near-optimal coding for an entire sequence of indices.
If the predictive distributions $q^{(i)}$ are correct, then the rate of the waveform coder will
be arbitrarily close to that of \eqref{q:genrate_est}.
Any mismatch in the conditional distribution results in an expected rate increase
that is specified by the Kulback-Leibler divergence.

The described WaveNet waveform coding scheme is close to optimal for the squared error
measure on samples with a $\mu$-law warped amplitude.  The cubic cell shape of scalar quantization
imposes a penalty
that is (asymptotically with increasing rate) maximally 1.5 dB in mean squared error distortion or,
equivalently, 0.25 bits per sample \cite{Lookabough:89}. As conditional distributions for
higher dimensionalities and the usage of analysis-by-synthesis suffer from high
complexity, the removal of this penalty is non-trivial.

The WaveNet waveform coder can be characterized by two rates. On the
one hand we can compute the average entropy rate of the estimated
conditional distribution:
\begin{align}
\bar{H}
 = - \frac{1}{|\A_0|} \sum_{i\in\A_0} \sum_{n\in\mathcal{N}} q^{(i)}_n \log_2 q^{(i)}_n ,
\label{q:estimated}
\end{align}
which provides an estimate of the true average entropy rate. On the other hand, we can
compute a lower bound on the real-world rate produced by the entropy coder:
\begin{align}
R = - \frac{1}{|\A_0|} \sum_{i\in\A_0}  \log_2 q^{(i)}_{n_i}.
\label{q:upperbound}
\end{align}
which forms an upper bound on the average entropy rate. If $R$ and $\bar{H}$
are close, we can be confident that the true average entropy rate is similar. The
measured rates will be discussed in section \ref{s:experiments}.

Finally, we discuss the rate required for the conditioning variables of a WaveNet waveform
coder. It was shown in \cite{Kleijn:07a} that under reasonable conditions \textit{the
  optimal rate for the  conditioning variables does not depend on the mean signal
  distortion}. This also applies to the WaveNet based waveform coder and implies that only
the resolution of the quantizer $Q$ has to be adjusted to vary the rate.

We did not include perceptual weighting in the current implementation of the waveform
WaveNet coder. However, pre- and postfiltering structures can be introduced to enhance
coder performance by exploiting perception.

\section {Experimental Results}
\label{s:experiments}
%\tcb{Half of page 3, most of page  4}
In this section, we evaluate the information rates, signal quality and speaker identifiability
produced by the WaveNet coding schemes. Listening examples are available online\footnote{\scriptsize\url{https://goo.gl/C14FFx}}.

\subsection{Experimental Setup}
The WaveNet system as described in \cite{WaveNet:2016a} was used to develop our proposed
WaveNet coders. At the encoder, we employed Codec 2 at 8 kHz and 2.4 kb/s.
As previously discussed, its decoded bit stream was used as WaveNet conditioning variables,
held constant over 10 ms intervals. The decoder then generated output samples at 16 kHz.

The training and test sets were derived from the Wall Street Journal speech corpus \cite{WSJdatabase}
with no overlapping speakers. The training set contained 32580 utterances by 123 speakers
and the testing set contained 2907 utterances by 8 speakers.

Quality was evaluated using POLQA and listening tests against a number of unmodified reference coders:
Codec~2 (2.4 kb/s), MELP (2.4 kb/s) \cite{McCree96}, Speex wideband (2.4 kb/s) \cite{speex},
ITU-T G.711 $\mu$-law at 16 kHz (128 kb/s), and ITU-T G.722.2 AMR-WB (23.05 kb/s).

Speaker identification performance of the parametric WaveNet coder was evaluated with listening tests and
a neural network based model \cite{Wan2018} trained on our dataset. As reference, we
trained a second parametric WaveNet coder with overlapping speakers in the training and test sets.

\subsection{Speech Information Rates}
\begin{figure}[t]
\vspace{-0.75em}
\centering
\includegraphics[width=1\columnwidth]{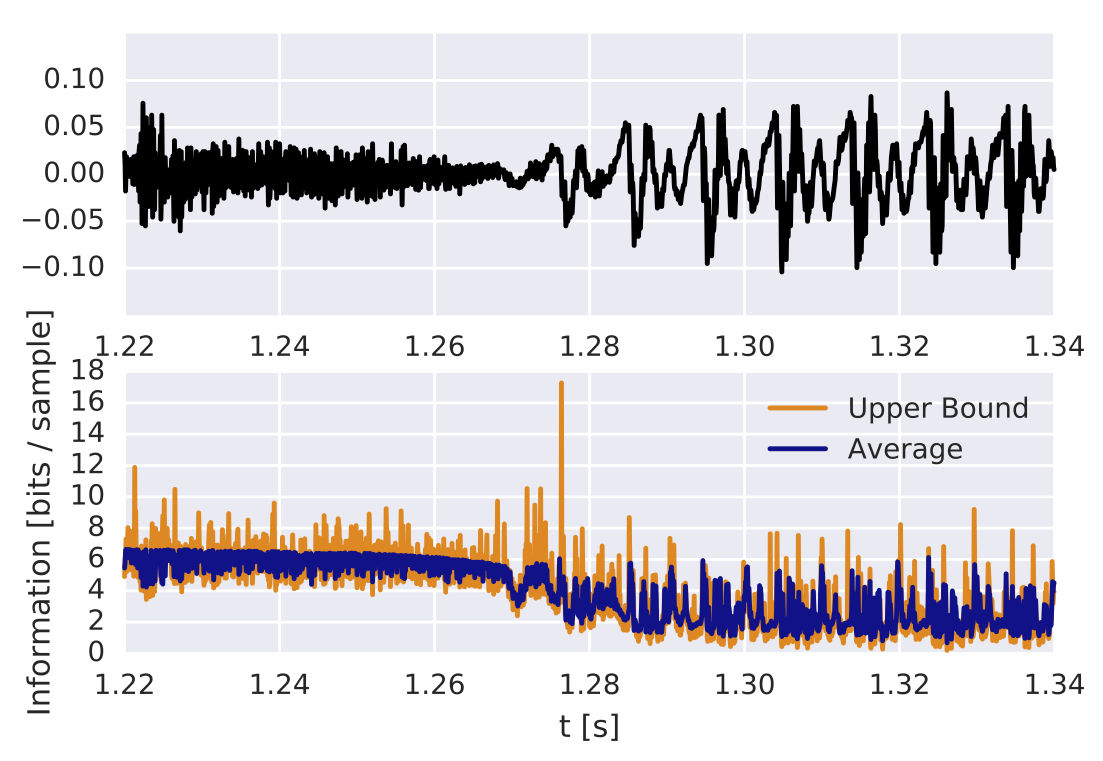}
\caption{Top: signal waveform. Bottom: instantaneous information computed as the terms in the sample average sums in \eqref{q:estimated} (blue) and
  \eqref{q:upperbound} (orange). \label{f:entropy}}
\end{figure}

We used \eqref{q:genrate_est} to estimate the rate of waveform
generation on the test dataset (removing silence segments). The result was a mean rate of 2.65 bits per \textit{sample},
or 42 kb/s at 16 kHz. Informal testing
indicates that this rate is largely independent of the rate of the conditioning parameters.
We also estimated the rates associated with \eqref{q:estimated} and \eqref{q:upperbound}
and obtained 2.61 and 2.62 bits per sample respectively, or about 42 kb/s at 16 kHz.
Since they are very similar, we can be confident that the true average entropy rate is similar.
Fig. \ref{f:entropy} shows a speech segment of 0.2 seconds and the
corresponding instantaneous information rate for the waveform coder. The early part of the
signal is a fricative, which is relatively unstructured, thus a higher rate is required. The latter part is a
voiced segment, where the rate required is low. The rate additionally varies with the pitch cycle. The highest
rate is not associated with the pitch pulse, implying that WaveNet predicts the pitch accurately, but the
waveform is noisier across certain segments of the pitch cycle.

\subsection{Quality Experiments}

\begin{table}[b]
  \centering
  \caption{POLQA mean opinion scores (MOS-LQO) for different coders operating at different rates (kb/s). WW: WaveNet
          waveform coder, WP: parametric WaveNet coder}
  \label{t:Qresults}
  \begin{tabular}{lcccccc}
    \hline
     & \textit{Codec 2} & \textit{MELP} & \textit{Speex} & \textit{AMR-WB} & \textit{WW} & \textit{WP}\\
    \hline
    Rate & 2.4 & 2.4 & 2.4 & 23  & 42  & 2.4\\
    MOS  & 2.7 & 2.9 & 2.2 & 4.6 & 4.7 & 2.9\\
    \hline
  \end{tabular}
\end{table}

The objective quality of the reference and two WaveNet coders was evaluated using POLQA
\cite{polqa} and the results are shown in Table
\ref{t:Qresults}. We noted that POLQA did not reflect informal listening impressions,
where the bandwidth extension and absence of the distortions typical of a
vocoder-based parametric coder was clearly heard.
This discrepancy was not unexpected as the parametric WaveNet
coder changes the signal waveform and the timing of the phones.

\begin{figure}
\centering
\includegraphics[width=1.0\columnwidth]{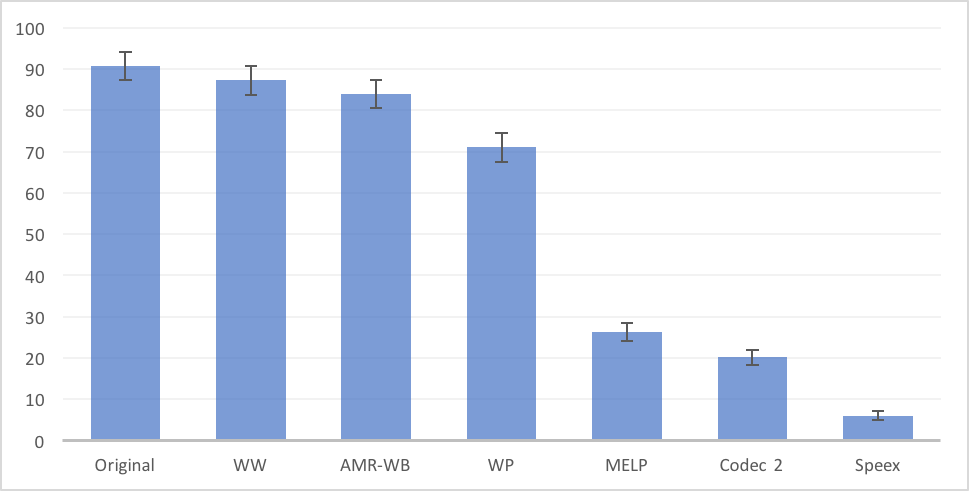}
%\vspace{0.3\columnwidth}
\caption{Subjective quality (MUSHRA scores). WW: WaveNet
         waveform coder, WP: parametric WaveNet coder. \label{f:squality}}
\end{figure}

A subjective MUSHRA-type listening test \cite{mushra} was performed where
21 participants evaluated 8 utterances. The $\mu$-law coder was omitted from the
test as it is identical to the WaveNet waveform coder.
The results are given in Fig. \ref{f:squality} where it can be seen that
two distinctive groups emerged:
a low-quality group consisting of Speex, Codec 2 and MELP, and a high-quality group
consisting of AMR-WB, the WaveNet waveform coder, and the parametric WaveNet coder. Thus,
the parametric WaveNet coder has a subjective quality similar to that of existing waveform coders
with the benefit of significantly lower rates.

\subsection{Speaker Identification Experiments}
An objective speaker identification test was performed using a neural network based
speaker identification model \cite{Wan2018}.
Two single-layer models were trained with $\mu$-law coded and parametric
WaveNet coded speech respectively. There were no overlapping speakers between the training and test data.
The training set contained 123 speakers and 3690 utterances and the same test set as before was used,
but now split into enrollment and verification sets with
overlapping speakers but non-overlapping utterances.
The verification equal error rate (EER) results were 8.4\% for the  $\mu$-law coded speech and 15.8\% for
WaveNet coded speech.  It is known that EER goes up after coding \cite{Dunn2001} and in this case,
it is expected that the spectral resolution
of the low rate coder has restricted speaker identifiability.

A listening test was also carried out. For this, a second parametric WaveNet coder was trained with overlapping speakers between the training and test data. We term this model as $\textit{W}_{\textit{w}}$ and the first model trained without overlapping speakers as $\textit{W}_{\textit{wo}}$. A triangle test was performed where 15 listeners listened to 16 trials, each with 3 different utterances by the same speaker. Two utterances were taken from one of \{$\textit{W}_{\textit{w}}$, $\textit{W}_{\textit{wo}}$\} and one utterance from the other model. The total utterances from each model were equal over all trials. The listeners had to indicate the utterance spoken by the different speaker. On average, they correctly identified the different speaker in 41\% of the trials. If the speakers are indistinguishable, the expected value is 33\%. This discrepancy is likely to diminish with more speakers in the training set.

These experiments indicate that the current coder would be more suitable for human-facing applications, e.g. conference calls.

%\begin{table}[b]
%  \centering
%  \caption{Speaker identity experiments for parametric WaveNet coder.}
%  \label{t:Sresults}
%  \begin{tabular}{lcc}
%    \hline
%                                       & false positive & false negative \\
%    \hline
%    Original speech      &       &  \\
%    WaveNet P coder      &     &    \\
%    \hline
%  \end{tabular}
%\end{table}

\section{Conclusions}
\label{s:conclusions}
Our results show that the high fidelity of the conditional probability distribution of the
speech waveform of WaveNet can be leveraged to create state-of-the-art speech and audio
coding systems. We demonstrated that the quality of our 2.4 kb/s parametric speech coder is similar
to that of waveform coders with much higher rates. We also showed how to build a waveform
WaveNet coder and briefly discussed a method for switching from the parametric coder to
the waveform coder based on a likelihood based quality measure for the parametric coder.

The computational cost of both training and running WaveNet is high compared to conventional
coders. The exception is the parametric WaveNet encoder, which is a conventional,
low complexity parametric encoder.

It is expected that performance can be improved further. For example, the conditioning
parameter set and their interpolation over time can be further refined and pre-/postfiltering can be introduced to the waveform coder to improve perceived performance. To increase
computational efficiency, it may be beneficial to study if the long-lag memory components
of the conditional probability distribution unnecessarily duplicates the information that is
also present in the bit stream.

\newpage
\bibliographystyle{IEEEtran}
\bibliography{AmbisonicsRefs,BeamFormerRefs,coding}
\end{document}